\documentclass{article}
\usepackage{arxiv}
\usepackage{cite}
\usepackage{amsmath,amssymb,amsfonts}
\usepackage{algorithmic}
\usepackage{graphicx}
\usepackage{textcomp}
\usepackage{array}
\usepackage[table,xcdraw]{xcolor}
\def\BibTeX{{\rm B\kern-.05em{\sc i\kern-.025em b}\kern-.08em
    T\kern-.1667em\lower.7ex\hbox{E}\kern-.125emX}}

\usepackage{pgfplots}
\usepackage{tikz}
\usetikzlibrary{positioning, shapes, arrows.meta}

\usepackage{makecell}
\usepackage{booktabs}
\usepackage{hyperref} 

\begin{document}
\title{Benchmarking Deep Learning Models on NVIDIA Jetson Nano for Real-Time Systems: An Empirical Investigation}


\author{Tushar Prasanna Swaminathan\\
        Electrical and Computer Engineering \\
        Lakehead University\\
        Thunder Bay, Canada \\
        \texttt{tswamina@lakeheadu.ca}\\
	\AND
        Christopher Silver\\
        Electrical and Computer Engineering \\
        Lakehead University\\
        Thunder Bay, Canada \\
        \texttt{crsilver@lakeheadu.ca}\\
        \And
	Thangarajah Akilan \\
	Department of Software Engineering\\
	Lakehead University\\
	Thunder Bay, ON P7B5E1 \\
	\texttt{takilan@lakeheadu.ca} \\
}

\maketitle

\begin{abstract}

The proliferation of complex deep learning (DL) models has revolutionized various applications, including computer vision-based solutions, prompting their integration into real-time systems. 
However, the resource-intensive nature of these models poses challenges for deployment on low-computational power and low-memory devices, like embedded and edge devices. 
This work empirically investigates the optimization of such complex DL models to analyze their functionality on an embedded device, particularly on the NVIDIA Jetson Nano. It evaluates the effectiveness of the optimized models in terms of their inference speed for image classification and video action detection. 
The experimental results reveal that, on average, optimized models exhibit a 16.11\% speed improvement over their non-optimized counterparts. This not only emphasizes the critical need to consider hardware constraints and environmental sustainability in model development and deployment but also underscores the pivotal role of model optimization in enabling the widespread deployment of AI-assisted technologies on resource-constrained computational systems. It also serves as proof that prioritizing hardware-specific model optimization leads to efficient and scalable solutions that substantially decrease energy consumption and carbon footprint.

\end{abstract}


\section{Introduction}

In the rapidly evolving landscape of DL, the size of models capable of handling complex tasks, viz. image classification, action recognition, object detection, and semantic segmentation has increased dramatically. This growth is driven by the abundant intricate information that these models need to extract and perceive from the input data.
Among them, ResNet~\cite{HeZRS15}, VGG~\cite{szegedy2014going}, Inception~\cite{Simonyan15}, and U-Net~\cite{azad2022medical} are examples of commonly utilized architectures. 
Meanwhile, the embedded and edge devices play a crucial role in performing real-time computation within operational environments. Deploying DL models on these devices eliminates the necessity for centralized cloud and data centers, leading to quicker and more reliable processing.  
Moreover, not everyone, be it an individual or an organization, can afford expensive high-performance computers (HPC) to deploy large DL models for practical applications. 
This problem is not only relevant to large DL models for large organizations' use cases, but there is also a notable surge in integrating DL technologies into mobile applications~\cite{liu2021smartphones}. This is especially relevant use case, as there will be a projected 6.38 billion smartphone users in the world~\cite{bankmycell}.  
Therefore, there is a need for optimizing existing models to run efficiently on devices with low computation and memory capacities.

This work extends beyond merely addressing the high computational demands on resource-constrained devices; it enables organizations to explore various modalities and combinations of deep learning models, thereby catalyzing breakthroughs in their respective fields that would otherwise be impractical on edge devices. For instance, Silver and Akilan \cite{Silver} demonstrated that integrating supervised and unsupervised models into a meta-model classifier yields superior performance compared to using each model individually. This approach is particularly beneficial for scenarios where collecting a comprehensive dataset is challenging. However, such implementations on edge devices remain impractical without model optimization to reduce complexity.
In this direction, this work pragmatically investigates DL model optimization for computer vision applications in-depth and conducts empirical analysis using the NVIDIA Jetson Nano. 

The rest of this paper is organized as follows: Section~\ref{sec-background} reviews related works and outlines the problem formulation. 
Section~\ref{sec-method} elaborates on methodology and experimental setup. Section~\ref{sec-results} discusses the experimental findings. 
Finally, Section~\ref{sec-conclusion} concludes the paper with future research directions.

\section{Background}\label{sec-background}

The literature indicates that while DL models are typically designed to produce high performance in their intended applications, they often struggle with real-time inference speed, a critical requirement for many practical applications. 
Consequently, models require meticulous optimization when deployed on edge devices, as the optimization strategies differ significantly from those used for specific hardware configurations. Without such tailored optimization, the models risk malfunctioning or failing entirely on the new hardware platform.
This lack of compatibility can constrain the deployment options and flexibility of deep learning (DL) solutions developed by engineers. Conversely, unoptimized models fail to efficiently utilize available hardware resources, such as central processing units (CPUs), graphics processing units (GPUs), or specialized accelerators like tensor processing units (TPUs). This inefficiency can result in elevated computational costs and extended inference times.
Hence, inefficient resource usage can lead to higher infrastructure costs, especially in cloud-based environments, where resources are typically billed based on usage~\cite{9459736}. 

\subsection{Cloud vs Edge Devices}

Numerous industrial applications are increasingly adopting Infrastructure as a Service (IaaS) provided by cloud vendors. However, the use of cloud computing services introduces several challenges, notably latency, which critically impacts real-time applications. For example, the frames captured by an autonomous vehicle’s camera must be processed immediately to prevent collisions with obstacles. Transmitting this video data over cloud networks often requires extensive bandwidth and memory, consequently leading to delays. Additionally, scalability issues arise, causing network access bottlenecks when data is simultaneously transmitted to the cloud from multiple sources. Privacy also remains a significant concern when transmitting sensitive data to the cloud.
In contrast, edge devices in operational environments, such as the aforementioned autonomous vehicle, can process data in real-time. They only transfer essential information, such as the results of object detection in a video stream, to the cloud server when necessary, optimizing bandwidth usage and reducing latency. Images and other critical user information are processed on the edge device without transferring them to the cloud, which protects from data being tapped or stolen~\cite{8763885}. 

\subsection{DL Model Optimization}

It is widely recognized that as deep learning (DL) architectures grow in size and complexity, they necessitate the use of highly computationally intensive devices, which consequently increases their carbon footprint. Model optimization, however, offers an effective solution to mitigate these environmental impacts. Studies demonstrate that models with higher resource requirements can be optimized and run on lower computation devices, resulting in a smaller carbon footprint~\cite{9407142}.
To address this, some researchers are concentrating on efficient hyperparameter optimization (HPO) that is suitable for deep learning (DL) models. Malik \textit{et al.} \cite{mallik2024priorband} propose the use of less computationally demanding proxy tasks during the training phase of DL models to conserve resources. Additionally, inefficient models tend to consume more power, thereby increasing energy consumption. In contrast, optimized models that demand fewer computational resources can significantly reduce the environmental impact associated with AI and DL deployments. Supporting such advancements, certain hardware platforms offer specialized features, such as tensor cores in GPUs, to boost model performance and integrate concepts like TinyML \cite{oliveira2024internet, 10.1145/3508391, sertic2022intelligent}.

There is also a growing demand for the development of new machine learning (ML) and neural network (NN) libraries that are optimized for use with field-programmable gate arrays (FPGAs) and application-specific integrated circuits (ASICs). For instance, Safaei~\textit{et~al.}~\cite{safaei2018system} introduce a ground-up approach to implement an extreme learning machine (ELM) classifier tailored to a system-on-a-chip (SoC) FPGA to gain computational acceleration, resulting in higher efficiency than general-purpose processors.
Hence, Wang~\textit{et al.}~\cite{9735379} introduce a hardware-friendly model optimization technique that is highly efficient on low-storage devices. Similarly, Sertic~\textit{et al.}~\cite{sertic2022intelligent} performed hardware-specific optimization of various object detection models to deploy them on Intel Neural Compute Stick 2 and Google Coral Edge TPU. 
Consequently, this paper benchmarks various optimized deep learning (DL) models on a NVIDIA Jetson Nano edge device, focusing on two computer vision tasks: image classification and human action recognition.

\section{The Methodology}\label{sec-method}

\subsection{Environmental Setup}\label{sec-experi-setup}

\begin{figure}[!ht]
    \centering
    \includegraphics[trim={0cm, 0.cm, 0.cm, 0.cm}, width=0.9\columnwidth, clip]{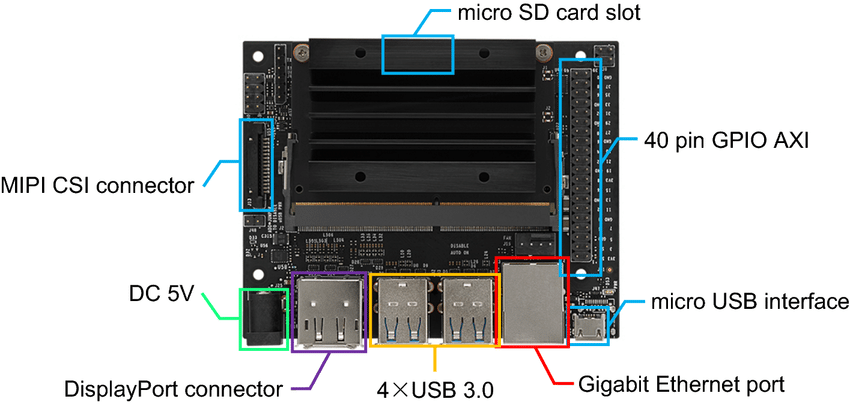}
    \caption{Layout of the NVIDIA Jetson Nano Developer Kit showcasing its key components and connectivity options \cite{JetsonNano}.}
    \label{fig:jetson}
\end{figure}

This section elaborates on the hardware and software tools used in this study.
The experimental study utilizes the NVIDIA Jetson Nano shown in Fig.~\ref{fig:jetson}, a compact and computationally powerful platform capable of parallel inference of multiple models. This platform is appropriate for various applications, including image classification, object detection, segmentation, tracking, and speech-to-text processing~\cite{MITTAL2019428}.
It consists of a CPU Quad-core ARM Cortex-A57 MP Core processor and GPU NVIDIA Maxwell architecture with 128 NVIDIA compute unified device architecture (CUDA) cores. 
The NVIDIA Maxwell architecture is suitable for deep learning processing tasks. The CUDA cores are the processing units responsible for performing computations on NVIDIA GPUs. 
The CUDA Toolkit provided by NVIDIA is a development environment for creating high-performance GPU-accelerated applications on the Jetson Nano~\cite{jetson-nano}.

The models are developed and tested using the PyTorch deep learning framework, which offers robust compatibility with \texttt{TensorRT} and facilitates straightforward conversion to \texttt{ONNX} format. \texttt{TensorRT}, an SDK by NVIDIA, is designed for real-time model inference, optimizing deep learning models to achieve low latency and high throughput. For model optimization and running inference with \texttt{TensorRT}, we use the \texttt{NVIDIA (L4T R32.7.1) PyTorch Container} for \texttt{Jetson JetPack 4.6.1}~\cite{container-nano}. This Docker container helps manage all compatible dependencies efficiently in a single environment, thereby preserving the integrity of external files.

\subsection{Experimental Setup}

The experiments are divided into two categories: image classification and human action recognition. For image classification, we utilize several pre-trained deep learning (DL) models from the native PyTorch library \cite{pytorch-vision}, including AlexNet, VGG, ResNet, SqueezeNet, DenseNet, ShuffleNet-V2, MobileNet-V2, and ResNet-V2. These models, known for their extensive use in various computer vision applications, have a high number of trainable parameters and large sizes, making them resource-intensive but ideal for benchmarking purposes in this study. All models are adapted for a binary image classification task.
For human action recognition, three custom models are developed: a 3D-CNN, an Autoencoder (AE), and a DenseNet, all designed to process video input streams. Training is conducted using the UCF50 dataset \cite{reddy2012recognizing}, which includes fifty different action categories. However, this study narrows its focus to a 4-way action classification to simplify the analysis.

\subsection{The Pipeline}\label{sec-optimization}

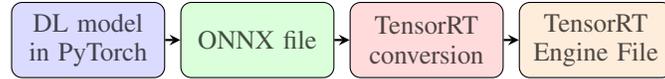
\begin{figure}[!tp]
    \centering

    \tikzset{
    block/.style={rectangle, rounded corners, minimum width=1cm, minimum height=1.0cm, text centered, text width=1.8cm, draw=black, fill=#1, fill opacity=0.7},
    arrow/.style={thick,->,>=stealth}
    }
    
    \begin{tikzpicture}[node distance=2.25cm]
    
    \node (dlmodel) [block=blue!20] {DL model in PyTorch};
    \node (onnx) [block=green!20, right of=dlmodel] {ONNX file};
    \node (conversion) [block=red!20, right of=onnx] {TensorRT conversion};
    \node (engfile) [block=orange!20, right of=conversion] {TensorRT Engine File};
    
    \draw [arrow] (dlmodel) -- (onnx);
    \draw [arrow] (onnx) -- (conversion);
    \draw [arrow] (conversion) -- (engfile);
    
    \end{tikzpicture}
    \caption{The PyTorch deep learning model optimization process for a NVIDIA Jetson Nano Edge Device using TensorRT.}
    \label{fig:conversion}
\end{figure}

Each model is optimized using \texttt{TensorRT} as mentioned above using the pipeline shown in Fig.\ref{fig:conversion}.

The conversion of a \texttt{PyTorch} deep learning model into a \texttt{TensorRT} engine comprises three main steps. The first step involves exporting the \texttt{PyTorch} model to an \texttt{ONNX} (Open Neural Network Exchange) file using the \texttt{torch.onnx.export()} function. This function traces the model's execution with a specified input tensor and exports the traced model to \texttt{ONNX} format. By default, the exported \texttt{ONNX} graph possesses fixed input sizes for all dimensions. To accommodate input tensors with dynamic batch sizes, it is crucial to set the first dimension as dynamic using the dynamic\_axes parameter of \texttt{torch.onnx.export()}. Exporting models to \texttt{ONNX} with dynamic shapes is essential for optimizing GPU memory usage and enhancing flexibility in processing varying input batch sizes.
From the \texttt{ONNX} file, the \texttt{TensorRT} engine is built using the \texttt{TensorRT} Python API. This involves several stages, including creating a network definition, importing the model through the \texttt{ONNX parser}, and building the \texttt{TensorRT} engine with a builder.

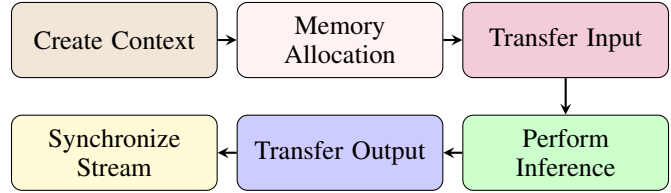
\begin{figure}[!t]
    \centering
    \tikzstyle{process} = [rectangle, minimum width=2.0cm, minimum height=1cm, text centered,text width=2.5cm, draw=black, rounded corners]
        \tikzstyle{arrow} = [thick,->,>=stealth]
        
        \begin{tikzpicture}[node distance=1.5cm]
        
        \node (step1) [process, fill=brown!20] {Create Context};
        \node (step2) [process, fill=pink!20, right of=step1, xshift=1.5cm] {Memory Allocation};
        \node (step3) [process, fill=purple!20, right of=step2, xshift=1.5cm] {Transfer Input};
        \node (step4) [process, fill=green!20, below of=step3] {Perform Inference};
        \node (step5) [process, fill=blue!20, left of=step4, xshift=-1.5cm] {Transfer Output};
        \node (step6) [process, fill=yellow!20, left of=step5, xshift=-1.5cm] {Synchronize Stream};
        
        \draw [arrow] (step1) -- (step2);
        \draw [arrow] (step2) -- (step3);
        \draw [arrow] (step3) -- (step4);
        \draw [arrow] (step4) -- (step5);
        \draw [arrow] (step5) -- (step6);   
        
        \end{tikzpicture}
    \caption{Inference process of \texttt{TensorRT} engine on NVIDIA Jetson Nano.}
    \label{fig:TensorRT-inference}
\end{figure}

The inference process with the \texttt{TensorRT} engine comprises six sub-steps (cf.~Fig.~\ref{fig:TensorRT-inference}): (i) creation of an inference execution context, (ii) memory allocation for input and output on the \texttt{CUDA} device, (iii) input data is transferred from the host into the input memory allocated on the \texttt{CUDA} device, (iv) \texttt{TensorRT} engine performs inference using the asynchronous execute API, (v) the output is transferred back into the host memory, and (vi) the stream used for data transfers and inference execution is synchronized to ensure the completion of all operations. This workflow allows for efficient deployment of \texttt{PyTorch} DL models on \texttt{TensorRT}-enabled platforms, leveraging the performance optimization capabilities of \texttt{TensorRT} for accelerated inference~\cite{10074837}. 
The \texttt{TensorRT} optimized file, i.e., \texttt{.trt} is generated through various techniques, like weight and activation precision calibration, layer and tensor fusion, kernel auto-tuning, dynamic tensor memory, and multi-stream execution methods for efficient inference by taking advantage of the NVIDIA Jetson GPU and its CUDA cores.

\begin{table*}[ht]
\caption{Performance comparison of base models (Pre-Opt) and their optimized counterparts (Post-Opt.) on a NVIDIA Jetson Nano before and after optimization. The Inference Time Speedup is Computed Against the Respective Pre-Opt Models. }
\label{tab:performance_comparison}
\centering
\setlength{\tabcolsep}{6pt} 
\begin{tabular}{|l|r|r|r|r|r|c|}
\hline\hline
\thead{\textbf{Model}} & \thead{\textbf{Pre-Opt.}\\ \textbf{Time} (s)} & \thead{\textbf{Post-Opt.}\\ \textbf{Time} (s)} & \thead{\textbf{Pre/Post}\\ \textbf{Time Ratio}} & \thead{\textbf{Trainable}\\ \textbf{Params}} & \thead{\textbf{FLOPS}} & \thead{\textbf{Inference Time}\\ \textbf{Speed up}} \\
\hline\hline
\rowcolor{blue!5}
\multicolumn{7}{|c|}{{Image Classification Models}} \\
\hline
AlexNet & 0.6638 & 0.1184 & 2.09 & 61,751,008 & 1,475,805,888 & $5.62\times$ \\
VGG & 2.5904 & 0.4285 & 6.05 & 143,667,240 & 19,590,954,654 & $6.05\times$ \\
ResNet & 1.0911 & 0.2233 & 4.88 & 25,557,032 & 3,883,453,952 & $4.88\times$ \\
SqueezeNet & 0.7966 & 0.1663 & 4.79 & 1,235,496 & 780,414,144 & $4.72\times$ \\
DenseNet & 1.0132 & 0.3682 & 2.16 & 7,978,856 & 2,845,996,288 & $2.72\times$ \\
ShuffleNet V2 & 4.7121 & 0.3463 & 13.61 & 2,278,604 & 298,632,608 & $13.6\times$ \\
MobileNet V2 & 5.0379 & 0.3003 & 16.77 & 3,504,872 & 300,356,480 & $16.7\times$ \\
ResNet V2 & 2.0964 & 0.2600 & 8.06 & 25,028,904 & 3,866,082,816 & $8.07\times$ \\
\hline
\rowcolor{blue!5}
\multicolumn{7}{|c|}{{Human Action Recognition Models}} \\
\hline
3D-CNN & 0.3431 & 0.0928 & 3.70 & 20,333,956 & 116,551,168 & $3.7\times$ \\
AutoEncoder (AE) & 0.7435 & 0.242 & 30.71 & 332,807 & 22,434,944 & $3.07\times$ \\
DenseNet & 3.1619 & 0.3718 & 8.50 & 109,386 & 27,681,924 & $8.5\times$ \\
\hline\hline
\end{tabular}%
\end{table*}

\section{Results and Discussions}\label{sec-results}

To test the optimized models, various parameters are taken into consideration, like floating point operations per second (FLOPS), and the inference time of the models before and after the optimization, noted hereafter as \textit{Pre-Opt} time and \textit{Post-Opt} time, respectively. 
Table~\ref{tab:performance_comparison} summarizes the models' details, including FLOPS and the number of trainable parameters as they are critical factors that affect inference time. 
FLOPS serves as a crucial metric for assessing the computational complexity of deep learning models. Typically, higher FLOPS implies a greater number of computations during inference, potentially leading to longer inference times.

For instance, models, like VGG with approximately 19 billion FLOPS exhibit a considerable decrease in inference time, achieving only a $6.05\times$ speedup after optimization.
Upon analyzing the numerical values in the table, a discernible trend emerges, wherein models with lower FLOPS tend to experience a more substantial decrease in inference time following optimization efforts. For example, Shuffle-V2 and MobileNet-V2 showcase speed up of $13.6\times$ and $16.7\times$, respectively, aligning with their low FLOPS values.

\begin{figure*}[!t]
    \centering
    \begin{tikzpicture}        
        \begin{axis}[
            ybar,
            ymin=0,
            ymax=20,
            bar width=0.7cm, 
            xtick=data,
            xticklabels={AlexNet, VGG, ResNet, SqueeNet, DenseNet, ShuffleNet-V2, MobileNet-V2, ResNet-V2, 3DCNN, AE, DenseNet}, 
            ylabel={Inference Time Speedup ($\times$)}, 
            width=0.9\textwidth, 
            height=8cm, 
            xticklabel style={rotate=45, anchor=north east}, 
            ymajorgrids, 
            grid style={dashed,gray!50}, 
            minor y tick num=1, 
            minor y tick style={gray!20}, 
        ]
            \addplot[fill=red!5] coordinates { 
                (1,5.62)
                (2,6.05)
                (3,4.88)
                (4,4.72)
                (5,2.72)
                (6,13.6)
                (7,16.5)
                (8,8.07)
                (9,3.7)
                (10,3.07)
                (11,8.5)
            };
        \end{axis}         
    \end{tikzpicture}    
    \caption{Inference time speedup of the optimized models on NVIDIA Jetson Nano compared to their non-optimized baseline counterparts.}
    \label{fig:barplot}
    \begin{flushleft}
        \vspace{-5.5cm}\noindent{\hspace{3.4cm}\footnotesize \rotatebox{90}{\texttt{1.48 MFLOPS}}\hspace{0.9cm}\rotatebox{90}{\texttt{19.59 MFLOPS}}\hspace{1.0cm}\rotatebox{90}{\texttt{3.88 MFLOPS}}\hspace{0.8cm}\rotatebox{90}{\texttt{0.78 MFLOPS}}\hspace{0.9cm}\rotatebox{90}{\texttt{2.85 MFLOPS}}\hspace{0.9cm}\rotatebox{90}{\texttt{0.30 MFLOPS}}\hspace{0.9cm}\rotatebox{90}{\texttt{0.30 MFLOPS}}\hspace{1.0cm}\rotatebox{90}{\texttt{3.87 MFLOPS}}\hspace{0.9cm}\rotatebox{90}{\texttt{0.12 MFLOPS}}\hspace{0.9cm}\rotatebox{90}{\texttt{0.02 MFLOPS}}\hspace{0.9cm}\rotatebox{90}{\texttt{0.03 MFLOPS}}}\vspace{3cm}
    \end{flushleft}
\end{figure*}
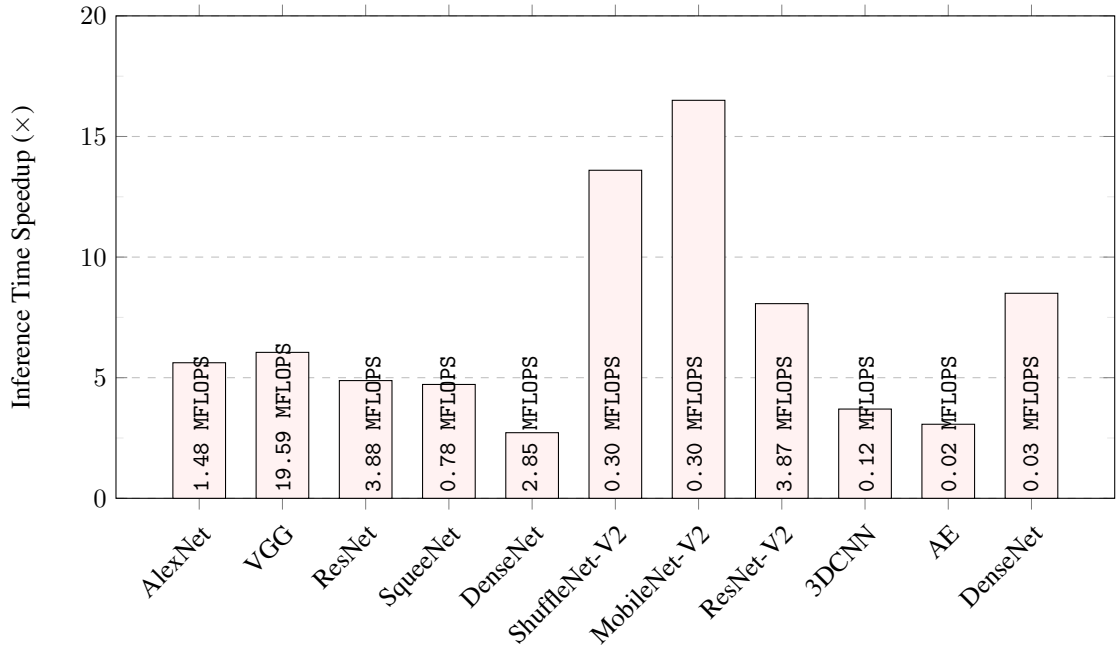

Despite the overarching trend, there exist exceptions within the dataset. Notably, ResNet-V2 possesses a significant number of FLOPS, yet its decrease in inference time is sped up at $8.07\times$ compared to other models with comparable FLOPS. 
This variation could stem from the unique architectural characteristics of ResNet and the effectiveness of optimization techniques applied during the model's development. Thus, it is clear that several factors contribute to the variations in the relationship between FLOPS and inference time, viz. optimization techniques, and network architecture, including the number of parameters, depth, and connectivity patterns that influence the model's computational complexity and subsequently the inference time. 
 
According to Fig.~\ref{fig:barplot}, there is a general trend where complex or large models show a minor reduction in inference time. Nevertheless, optimization positively affects the inference time across the models. However, the video-based action recognition models, specifically DenseNet and AutoEncoder (AE), do not conform to this trend. This divergence may be attributed to their architectural designs and the inefficiency of layer optimization, stemming from their status as custom-created rather than pre-trained models, which typically undergo extensive optimization.

\section{Conclusion and Future Work}\label{sec-conclusion}

Considering the environmental impact of hardware devices, sustainable growth in AI deployment involves optimizing models rather than solely relying on hardware upgrades. By reducing computational complexity and resource requirements, optimized models contribute to minimizing energy consumption and carbon footprint, aligning with the goals of sustainability and environmental conservation. 
The results of this empirical study demonstrate the significant impact of optimization techniques on various DL-driven computer vision models. On average the optimized models are $7.011\times$ faster than their non-optimized baseline counterparts. By leveraging tools, like \texttt{TensorRT}, the inference time of complex models is substantially reduced, making them more suitable for time-sensitive applications. The relationship between FLOPS and inference time highlights the importance of optimization in managing computational complexity and improving performance. Despite variations among models, the overarching trend showcases the effectiveness of optimization in decreasing inference time. 
While this work primarily focuses on optimization using \texttt{TensorRT}, future research can investigate other optimization techniques, such as quantization-aware training (QAT), and network pruning to further improve model efficiency and scalability.
In conclusion, optimizing deep learning (DL) models for edge and embedded devices, such as the NVIDIA Jetson Nano, is essential for enabling efficient and scalable AI solutions. These optimization techniques address the challenges associated with cloud-based model deployment and are pivotal for fostering sustainable growth in AI and enhancing edge computing capabilities in microcontrollers and single-board computers.
For further details and access to the source code of this work, the reader can refer to the \href{https://github.com/xTOTODILEx/Deep-learning-model-Optimization}{project repository}.


\bibliographystyle{ieeetr}  
\bibliography{references}

\end{document}